\documentclass{aa}
\usepackage{times}
\usepackage{graphics}
\usepackage{astronmar}

\begin{document}
\thesaurus{06(02.08.1; 08.02.1; 08.15.1; 08.18.1)} 
\title{The dynamical tide in a rotating 10~$\mathrm{M}_\odot$ main sequence star}
\subtitle{a study of g- and r-mode resonances}
\titlerunning{g- and r-mode resonances in a 10~$\mathrm{M}_\odot$ star}
\author{M.G.~Witte\thanks{marnix@astro.uva.nl} \and G.J.~Savonije\thanks{gertjan@astro.uva.nl}}

\institute{Astronomical Institute `Anton Pannekoek', University of
  Amsterdam, Kruislaan 403,  1098 SJ Amsterdam, The Netherlands}

\date{Received 19 July 1998 / Accepted <date>}
\maketitle

\begin{abstract} We study the linear, but fully non-adiabatic tidal response of
a uniformly rotating, somewhat evolved ($X_\mathrm{c}=0.4$), 10~$\mathrm{M}_\odot$ main sequence
star to the dominant $l=2$ components of its binary companion's tidal potential.
This is done numerically with a 2D implicit finite difference scheme.  We assume
the spin vector of the 10~$\mathrm{M}_\odot$ star to be aligned perpendicular to the
orbital plane and calculate the frequency $\bar{\sigma}$ and width of the resonances
with the prograde and retrograde gravity (g) modes as well as the resonances
with quasi-toroidal rotational (r) modes for varying rotation rates $\Omega_\mathrm{s}$ of
the main sequence star.  For all applied forcing frequencies we determine the
rate of tidal energy and angular momentum exchange with the companion.  In a
rotating star tidal energy is transferred from $l=2$ g-modes to g-modes of
higher spherical degree ($l=4,6,8,\ldots$) by the Coriolis force.  These latter
modes have shorter wavelength and are damped more heavily, so that the $l=2$
resonant tidal interaction tends to be reduced for large rotation rates $\Omega_\mathrm{s}$.
On the other hand, the density of potential resonances (a broad $l$ spectrum)
increases.  We find several inertially excited unstable $l>4$ g-modes, but not
more than one (retrograde) unstable $l=2$ g-mode and that only for rapid
rotation.  Our numerical results can be applied to study the tidal evolution of
eccentric binaries containing early type B-star components.

\end{abstract}
  

\section{Introduction} Zahn \cite*{Z77} initiated the study of radiative damping
of the dynamical tide as a viable mechanism for effective tidal interaction in
early type close binary systems.  Savonije \&\ Papaloizou \cite*{SP84} were the
first to perform fully non-adiabatic calculations of the dynamical tide and to
study the interplay between stellar- and tidal evolution which appears crucial
in understanding the effects of tides in early type stars.  More recently,
Papaloizou \&\ Savonije \cite*{PS97} and Savonije \&\ Papaloizou \cite*{SP97}
(from now on SP97) studied the effects of rotation on the dynamical tide.  To
this end a 2D implicit code was developed for which the effects of the Coriolis
force on the non-radial oscillations are taken fully into account.  These
earlier studies of rotational effects were based on a chemically homogeneous
20~$\mathrm{M}_\odot$ star and were focussed on the low-frequency inertial regime, i.e.\ to
forcing frequencies $\bar{\sigma} < 2 \Omega_\mathrm{s}$, where $\bar{\sigma}$ is the forcing frequency in
the frame corotating with the star at a rate $\Omega_\mathrm{s}$.  The present study extends
these calculations to a somewhat evolved (with core hydrogen abundance
$X_\mathrm{c}=0.4$) 10~$\mathrm{M}_\odot$ star whereby the forcing frequency $\bar{\sigma}$ runs from low
(high radial order g-modes) to high frequencies (up to g$_1$).  The motivation
for this work is that we intend to apply the calculated energy and angular
momentum exchange rates with the companion to study the tidal evolution of
eccentric early type binary systems.  Recently several new studies of tidal
evolution in eccentric early type binaries (with a compact companion) have
appeared \cite[e.g.]{KG96,L97}.  However, these studies have not considered the
important effect of rotational (r) modes on the tidal exchange of angular
momentum in eccentric binary systems.  We anticipate interesting tidal effects
in eccentric binaries by the counteracting effects of resonant prograde g-modes
and retrograde r-modes when the early type component is rotating near its
`pseudo' synchronous rate at periastron.  This will be studied in a following
paper.

\section{Basic equations} We consider a uniformly rotating, early type star with
mass $M_\mathrm{s}$ and radius $R_\mathrm{s}$ in a close binary with circular orbit with angular
velocity $\omega$ and orbital separation $a$.  We assume the stellar angular
velocity of rotation $\vec{\Omega}_\mathrm{s}$ to be much smaller than its
break-up speed, i.e.\ $(\Omega_\mathrm{s}/\Omega_\mathrm{c})^2\ll 1$, with $\Omega_\mathrm{c}^2=GM_\mathrm{s}/R_\mathrm{s}^3$, so that
effects of centrifugal distortion ($\propto \Omega_\mathrm{s}^2$) may be neglected in first
approximation.  We wish to study the response of this uniformly rotating star to
a perturbing time-dependent tidal force.  The Coriolis acceleration is
proportional to $\Omega_\mathrm{s}$ and we take its effect on the tidally induced motions in
the star fully into account.  We use spherical coordinates
$(r,\vartheta,\varphi)$, with origin at the stellar centre, whereby
$\vartheta=0$ corresponds to its rotation axis which we assume to be parallel to
the orbital angular momentum vector.  We take the coordinates to be
non-rotating.

As is well known, in a non-rotating star the solutions of the linearized
non-radial stellar oscillation equations can be expressed in terms of spherical
harmonics, i.e.\ the spatial part of each mode can be fully separated into r-,
$\vartheta$- and $\varphi$-factors \cite[e.g.]{LW58} $U(r,\vartheta,\varphi) =
u(r) \,P^m_l(\cos \vartheta)\, \mathrm{e}^{- \mathrm{i} m\varphi}$, where $P^m_l$
represents the associated Legendre polynomials for $l$ and $m$.

The introduction of the Coriolis force, however, destroys the full separability
of the oscillation equations, it only being retained for the
$\varphi$-dependence.  It turns out \cite[e.g.]{BG78} that two independent sets
of approximately spheroidal oscillation modes exist:  modes in which the density
perturbation is \emph{even} with respect to reflection in the equatorial plane,
which have $l-|m|$ even valued, and modes with \emph{odd} symmetry for the
density, having $l- |m|$ odd valued.  In addition, for each $l$, there is a set
of quasi-toroidal r-modes \cite[e.g.]{PP78} which couple with the spheroidal
modes of $l \pm 1$.

Let us denote perturbed Eulerian quantities like pressure $P'$, density
$\rho'$, temperature $T'$ and energy flux $\vec{F}'$ with a prime.
The linearized hydrodynamic equations governing the non-adiabatic response of
the uniformly rotating star to the perturbing potential $\Phi_\mathrm{T}$ may then be
written
\[ 
\left[ \left(\frac{\partial }{\partial t} + \Omega_\mathrm{s}
    \frac{\partial }{\partial \varphi}\right)
  v_i\right] {\hat{\vec{e}}_i }+ 2 \Omega_\mathrm{s} \vec{k} \times \vec{v}'
\]
\begin{equation} 
  =-\frac{1}{\rho}
  \nabla P' + \frac{\rho'}{\rho^2} \nabla P - \nabla \Phi_\mathrm{T},
  \label{eqmot} 
\end{equation}
\begin{equation}  
  \left(\frac{\partial }{\partial t} + \Omega_\mathrm{s} \frac{\partial }{\partial \varphi}\right) \rho' +
  \nabla\cdot\left(\rho \vec{v}' \right) =0, 
  \label{eqcont} 
\end{equation}
\begin{equation}  
  \left(\frac{\partial }{\partial t} + \Omega_\mathrm{s} \frac{\partial }{\partial \varphi}\right) \left[ S' +
    \vec{v}'\cdot \nabla S \right]=-\frac{1}{\rho T} \nabla\cdot\vec{F}', 
  \label{eqe}
\end{equation}
\begin{equation} 
  \frac{\vec{F}'}{F}=\left({{\mathrm{d} T}\over {\mathrm{d} r}}\right)^{-1} \left[ \left(\frac{3
        T'}{T} -\frac{\rho'}{\rho} -\frac{\kappa'}{\kappa} \right) \nabla
    T + \nabla T' \right],
  \label{eqf} 
\end{equation}
where $\hat{\vec{e}}_i$ are the unit vectors of our spherical coordinate system,
\vec{k} is the unit vector along the rotation axis, $\vec{v}'$ denotes the
velocity perturbation, $\kappa$ the opacity of stellar material and $S$ its
specific entropy.  These perturbation equations represent, respectively,
conservation of momentum, conservation of mass and conservation of energy, while
the last equation describes the radiative diffusion of the perturbed energy
flux.  For simplicity we adopt the Cowling \cite*{C41} approximation, i.e.\ we
neglect perturbations to the gravitational potential caused by the stellar
distortion.  We also neglect perturbations of the nuclear energy sources and of
convection.

For a circular orbit (with orbital angular speed $\omega$) the companion's
perturbing potential can be expanded as the real part of \cite[e.g.]{MF53}:
\[  \Phi_\mathrm{T}(r,\vartheta,\varphi,t)= -\frac{G M_\mathrm{p}}{a} \sum_{l=0}^{\infty}
\sum_{m=0} ^l \epsilon_m \frac{\left( l-m\right)!}  {\left( l+m\right)!}
\left(\frac{r}{a}\right)^l 
\]
\begin{equation} 
  \cdot P^{m}_{l}(\cos \vartheta)\, P^{m}_{l}(\cos
  \frac{\pi}{2})\, \mathrm{e}^{\mathrm{i}  m \left(\omega t- \varphi \right)} 
  \label{eq:pot} 
\end{equation}
where $M_\mathrm{p}$ is the companion's mass, $a$ the orbital separation,
$P^{m}_{l}(\cos \vartheta)$ the associated Legendre polynomial and
$\epsilon_m=1$ for $m=0$ and 2 for $m>0$.  We will consider only the dominant
$l=2$ components of the tidal forcing.  Adopting the same azimuthal $m$ symmetry
and time dependence for the perturbed quantities as the forcing potential, the
perturbed velocity can be expressed as $\vec{v}'= \mathrm{i} \bar{\sigma} \vec{\xi}$, where
$\bar{\sigma}=m (\omega-\Omega_\mathrm{s})$ is the forcing frequency felt by a mass element in the
uniformly rotating star and $\vec{\xi}$ is the displacement vector.

The perturbations can be written as e.g.\ $
\xi_r(r,\vartheta,\varphi,t)=\widehat{\xi_r}(r,\vartheta)\,\mathrm{e}^{\mathrm{i}
m(\omega t-\varphi)} $ where $\xi_r$ is the radial component of the displacement
vector, while $\widehat{\xi_r}(r,\vartheta)$ is assumed complex to describe the
azimuthal phase shift with respect to the forcing potential~(\ref{eq:pot})
induced by any occurring dissipation, e.g.\ turbulent viscosity or radiative
damping, see energy equation below.

The current equations contain extra terms compared to those in SP97 because of
the occurring mean-molecular weight ($\mu_a$)-gradients near the edge of the
convective core.  We assume diffusive mixing to be negligible on the
(oscillation) timescales under consideration, so that the Lagrangian variation
of the mean molecular weight $\delta \mu_a=0$ or $ \frac{\mu'_a}{\mu_a}=- 
{{\mathrm{d} \ln \mu_a}\over {\mathrm{d} r}} \xi_{r}$.  We can thus use
\begin{equation} 
\frac{P'}{P}=\chi_{\rho} \left(\frac{\rho'}{\rho}\right) + \chi_T \left(\frac{T'}{T}\right) -\chi_{\mu}
{{\mathrm{d} \ln \mu_a}\over {\mathrm{d} r}} \xi_r 
\label{eqst} 
\end{equation} 
to eliminate the pressure perturbation, where $\chi_{\rho}=\frac{\partial\ln P}{\partial \ln 
\rho}$,
$\chi_{T}=\frac{\partial\ln P}{\partial \ln T}$ and $\chi_{\mu}=\frac{\partial\ln P}{\partial \ln \mu_a}$ follow 
from
the equation of state.

Writing for simplicity from now on $\xi_r$ for 
$\widehat{\xi_r}(r,\vartheta)$, 
etc.,
while dividing out the factor $\mathrm{e}^{\mathrm{i} m (\omega t-\varphi)}$, 
Eqs.~\ref{eqmot}--\ref{eqf} yield the seven scalar 
Eqs.~\ref{eqco}--\ref{eqFt} given below.

First of all we write out the perturbed equation of continuity
\begin{equation}
  \frac{\rho'}{\rho}=-\frac{1}{r^2\rho}\frac{\partial }{\partial r}\left(r^2\rho\xi_r\right) 
  -\frac{1}{r \sin{\vartheta}} \frac{\partial }{\partial \vartheta} \left(\sin{\vartheta} \, 
    \xi_{\vartheta}\right) 
  + \frac{\mathrm{i} m}{r \sin{\vartheta}} \xi_{\varphi}. 
  \label{eqco}
\end{equation}
We can use (\ref{eqco}) to eliminate the term $\frac{\partial\xi_r}{\partial r}$ introduced by
the radial derivative of the pressure perturbation (through~(\ref{eqst})) from
the radial equation of motion, so that the latter equation can be expressed
(after adding viscous terms to introduce turbulent damping in convective
regions, see SP97) as
\[ 
\left[\rho \bar{\sigma}^2 -\frac{P \chi_{\mu}}{\rho r^2}
  {{\mathrm{d} \ln \mu_a}\over {\mathrm{d} r}}
  {{\mathrm{d} (\rho r^2)}\over {\mathrm{d} r}} + P
  {{\mathrm{d} (\chi_{\mu} {{\mathrm{d} \ln \mu_a}\over {\mathrm{d} r}})}\over {\mathrm{d} r}}
\right.
\]
\[
\left.
 +\chi_{\mu}
  {{\mathrm{d} P}\over {\mathrm{d} r}}
  {{\mathrm{d} \ln \mu_a}\over {\mathrm{d} r}} \right] \xi_r
  + \left[2 \mathrm{i} \rho \bar{\sigma} \Omega_\mathrm{s} \sin{\vartheta} + \mathrm{i}  m \frac{P \chi_{\mu}}{r 
      \sin \vartheta} {{\mathrm{d} \ln \mu_a}\over {\mathrm{d} r}} \right] \xi_{\varphi}
\]
\[
+  \mathrm{i} \bar{\sigma} \frac{\rho \zeta}{r^2} \left[ \left(1-\mu^2\right) 
    \frac{\partial^2}{\partial \mu^2}
    \xi_r -2 \mu \frac{\partial}{\partial \mu} \xi_r -\frac{4}{1-\mu^2} \xi_r \right]
\]
\[
  +\frac{\mathrm{i} \bar{\sigma}}{r^2} \frac{\partial}{\partial r}\left(\rho\zeta r^2 \frac{\partial\xi_r}{\partial r} 
  \right)
  - \frac{P \chi_{\mu}}{r \sin \vartheta} {{\mathrm{d} \ln \mu_a}\over {\mathrm{d} r}} \frac{\partial}{\partial \vartheta}(\sin \vartheta 
  \,\xi_{\vartheta})
\]
\[
 + \left[
    {{\mathrm{d} P}\over {\mathrm{d} r}} -P \chi_{\mu} {{\mathrm{d} \ln \mu_a}\over {\mathrm{d} r}} 
    -P {{\mathrm{d} \chi_{\rho}}\over {\mathrm{d} r}} -\chi_{\rho} {{\mathrm{d} P}\over {\mathrm{d} r}} \right] 
  \left(\frac{\rho'}{\rho}\right)
\]
\[
  -P \chi_{\rho} \frac{\partial}{\partial r} \left(\frac{\rho'}{\rho}\right) 
  -\left[ P {{\mathrm{d} \chi_{T}}\over {\mathrm{d} r}} + \chi_{T} {{\mathrm{d} P}\over {\mathrm{d} r}} \right] 
  \left(\frac{T'}{T}\right)
\]
\begin{equation}
  - P \chi_{T} \frac{\partial}{\partial r}
  \left(\frac{T'}{T}\right) 
  = - \rho \frac{\partial\Phi_\mathrm{T}}{\partial r}
  \label{eqmr} 
\end{equation}
where $\mu=\cos \vartheta$ and $\zeta$ is the coefficient of turbulent viscosity
defined below.  The $\vartheta$-equation of motion becomes
\[
\rho \bar{\sigma}^2 \xi_{\vartheta} + \mathrm{i} \bar{\sigma} \frac{\rho \zeta}{r^2} 
\left[\left(1-\mu^2 \right) \frac{\partial^2}{\partial \mu^2}
  \xi_{\vartheta} -4 \mu \frac{\partial} {\partial \mu}\xi_{\vartheta} -\frac{5-2 \mu^2}{1- 
\mu^2}
  \xi_{\vartheta} \right]
\]
\[
+\frac{\mathrm{i} \bar{\sigma}}{r^2} \frac{\partial}{\partial r} \left(\rho \zeta r^2 
  \frac{\partial\xi_{\vartheta}}{\partial r}
\right)
+\left[2 \mathrm{i} \rho \bar{\sigma} \Omega_\mathrm{s} \cos{\vartheta} \right] \xi_{\varphi}
\]
\[
-\left(\frac{P \chi_{\rho}}{r}\right) \frac{\partial }{\partial \vartheta} \left(\frac{\rho'}{\rho}\right) - 
\left(\frac{P \chi_T}{r}\right)
\frac{\partial }{\partial \vartheta} \left(\frac{T'}{T}\right) 
\]
\begin{equation}
  + \frac{P \chi_{\mu}}{r} {{\mathrm{d} \ln \mu_a}\over {\mathrm{d} r}} \frac{\partial\xi_r}{\partial \vartheta}=
  -\frac{\rho}{r} \frac{\partial\Phi_\mathrm{T}}{\partial \vartheta}. 
  \label{eqmt} 
\end{equation}
The $\varphi$-equation of motion can be expressed as 
\[
\rho \bar{\sigma}^2 \xi_{\varphi}
-\left[ 2 \mathrm{i} \rho \bar{\sigma} \Omega_\mathrm{s} \sin{\vartheta} +\mathrm{i}  m \frac{P 
    \chi_{\mu}}{r\,
    \sin \vartheta} {{\mathrm{d} \ln \mu_a}\over {\mathrm{d} r}} \right] \xi_r
\]
\[
- \left[2 \mathrm{i}  \rho \bar{\sigma} \Omega_\mathrm{s} 
  \cos{\vartheta}
\right] \xi_{\vartheta} + \left[\frac{\mathrm{i} m P \chi_{\rho}}{r 
    \sin{\vartheta}}\right] \left(\frac{\rho'}{\rho}\right)
\]
\begin{equation} 
  + \left[\frac{\mathrm{i} m P \chi_{T}}{r \sin{\vartheta}}\right] \left(\frac{T'}{T}\right) 
  =\frac{\mathrm{i} m \rho}{r\, \sin \vartheta} \Phi_\mathrm{T}. 
  \label{eqmp} 
\end{equation}

By applying the thermodynamic relation
\[ \delta S=S'+\vec{\xi}\cdot\nabla S= \frac{P}{\rho T}\frac{1}{\Gamma_3-1}
\left(\frac{\delta P}{P} -\Gamma_1 \frac{\delta \rho}{\rho} \right) \]
where the symbol $\delta$ denotes a Lagrangian perturbation and $\Gamma_j$
the adiabatic exponents of Chandrasekhar, the perturbed energy equation can
be expressed as
\[
\left[{{\mathrm{d} \ln P}\over {\mathrm{d} r}}-\Gamma_1 {{\mathrm{d} \ln \rho}\over {\mathrm{d} r}}-\chi_{\mu} {{\mathrm{d} \ln \mu_a}\over {\mathrm{d} r}} \right] \xi_r +
\left[\chi_{\rho} - \Gamma_1 \right]\left(\frac{\rho'}{\rho}\right)
\]
\[
+ \left[\chi_T + \mathrm{i}  \eta
  \left(\frac{m}{r \sin \vartheta}\right)^2 \left({{\mathrm{d} \ln T}\over {\mathrm{d} r}}\right)^{-1}\right] \left(\frac{T'}{T}\right)  
\]
\[
-\mathrm{i} \eta \left[ \frac{1}{r^2}\frac{\partial}{\partial r} \left(r^2\frac{F'_r}{F}\right) + {{\mathrm{d} \ln F}\over {\mathrm{d} r}}
 \left(\frac{F'_r}{F}\right) \right]
\]
\begin{equation}    
  + \mathrm{i} \eta \left[\frac{\sin
      \vartheta}{r}\frac{\partial}{\partial \mu}\left(\frac{F'_{\vartheta}}{F}\right) -\frac{\cos 
      \vartheta}{r \sin
      \vartheta} \left(\frac{F'_{\vartheta}}{F}\right) \right] = 0 
  \label{eqen} 
\end{equation}
where $\eta=\left(\Gamma_3-1\right)\frac{F}{\bar{\sigma} P} $ is a local characteristic
radiative diffusion length in the star, with $F$ the unperturbed (radial) energy
flux.  We have eliminated $F'_{\varphi}$ with help of the $\varphi$-component of
the radiative flux equation.  In the stellar interior $\eta \simeq 0$ which
corresponds to almost adiabatic response.  However, even for `high' frequencies
$\bar{\sigma} /\Omega_\mathrm{c} \approx 1$ the diffusion length $\eta$ becomes comparable to the
scale height when the stellar surface is approached.  The associated radiative
energy losses give rise to damping of the tidally excited oscillations whereby
the resulting phase lag with the companion generates a torque.

The perturbed radial radiative energy flux is given by
\[
  \left(\frac{F'_r}{F}\right) 
  =\left({{\mathrm{d} \ln T}\over {\mathrm{d} r}}\right)^{-1} \frac{\partial}{\partial r} \left(\frac{T'}{T}\right)  - \left(\kappa_T -4\right)
 \left(\frac{T'}{T}\right)
\]
\begin{equation} 
  -\left(\kappa_{\rho} +1\right)  \left(\frac{\rho'}{\rho}\right) + \kappa_X \frac{\partial\ln X}{\partial r} \xi_r
  \label{eqFr}
\end{equation}
where $\kappa_{\rho}=\frac{\partial\ln \kappa}{\partial \ln \rho}$, \,$\kappa_{T}=\frac{\partial\ln
  \kappa}{\partial \ln T} $ and $\kappa_{X}=\frac{\partial\ln \kappa}{\partial \ln X}$, with $X$ for 
the
hydrogen abundance.  

Finally the $\vartheta$-component of the perturbed flux follows as
\begin{equation} 
  \left(\frac{F'_{\vartheta}}{F}\right) = \left({{\mathrm{d} \ln T}\over {\mathrm{d} r}}\right)^{-1} \frac{1}{r}
  \frac{\partial}{\partial \vartheta} \left(\frac{T'}{T}\right). 
  \label{eqFt} 
\end{equation}

\subsection{Boundary conditions}
The differential equations are supplemented by the following boundary
conditions: at the stellar centre we require $\xi_r$ and $F'_r$ to vanish,
while at the stellar surface, we require the Lagrangian pressure
perturbations to vanish
\begin{equation} 
  \frac{P'}{P}+ {{\mathrm{d} \ln P}\over {\mathrm{d} r}} \xi_r=0 
  \label{bcs1} 
\end{equation}
and the temperature and flux perturbations to fulfill 
Stefan--Boltzmann's law  
\begin{equation} \frac{F'_r}{F}=\left(\frac{2}{r}+4 
{{\mathrm{d} \ln T}\over {\mathrm{d} r}}\right) \xi_r + 4 \left(\frac{T'}{T}\right). 
  \label{bcs2} 
\end{equation}
Furthermore, $\xi_{\vartheta}$ and $F'_{\vartheta}$ must vanish on the rotation
axis while in view of the symmetry of the tidal force we adopt mirror symmetry
about the equatorial plane, i.e., for $\vartheta=\pi/2$ we also require
$\xi_{\vartheta}$ and $F'_{\vartheta}$ to vanish.

\subsection{The unperturbed stellar model} \label{sec:model}
\begin{figure}
\includegraphics{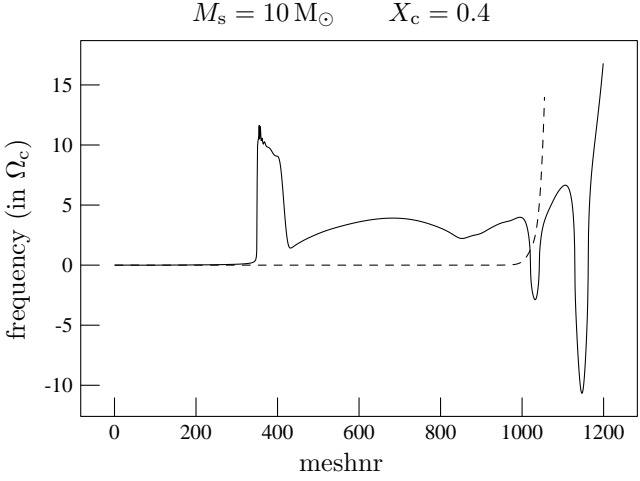} \caption[]{Characteristics of the stellar model:
continuous curve represents the Brunt-V\"{a}is\"{a}l\"{a} frequency
$\nu_\mathrm{BV}= \mathrm{sign}\mathcal{A}\sqrt{|\mathcal{A}|}$ in units of the
stellar break-up speed$\Omega_\mathrm{c}$ as a function of radial mesh number.  The dashed
curve shows the forcing frequency $\bar{\sigma}$ for which the corresponding
oscillation period equals the local thermal timescale.}  \label{fIM}
\end{figure}
A recent version \cite{PT95} of the stellar evolution code
developed by Eggleton~\cite*{E72} was used to construct the unperturbed stellar
input model.  The model represents a somewhat evolved main-sequence star of
10~$\mathrm{M}_\odot$ with core hydrogen abundance of $X=0.4$ and $Z=0.02$.  
The mass in
the convective core is approximately 2.1~$\mathrm{M}_\odot$.  
The model
comprises 1200 (radial) zones and was constructed with the OPAL opacities
\cite{IR96}.  The stellar radius equals $R_\mathrm{s}=3.825 \times 10^{11}$ cm, while the
effective temperature $T_\mathrm{eff}=2.314 \times 10^4$~K and the stellar
moment of inertia $I_\mathrm{s}=1.56\times 10^{56} \mbox{ g cm$^2$}$.  The
break-up angular speed equals $\Omega_\mathrm{c}=1.54 \times 10^{-4} \mbox{ s$^{-1}$}$.  The
Brunt-V\"{a}is\"{a}l\"{a} frequency $\mathcal{A}= {1\over
\rho}{{\mathrm{d} P}\over {\mathrm{d} r}}\left({1\over \rho}{{\mathrm{d} \rho}\over {\mathrm{d} r}} 
-{1\over \Gamma_1 P}{{\mathrm{d} P}\over {\mathrm{d} r}}\right)$ given in units of 
$\Omega_\mathrm{c}$, is plotted in Fig.~\ref{fIM}.
The Brunt-V\"{a}is\"{a}l\"{a} frequency attains large values in the region where
the convective core has retreated during the evolution and a composition
gradient is formed (the `$\mu_a$-gradient zone').  It can also be seen that
there are two shallow convective shells near the stellar surface.  
The dashed curve shows the
`thermal frequency' $\nu_\mathrm{th}= 2 \pi /\tau_\mathrm{th}$, where $
\tau_\mathrm{th}=\frac{\rho \kappa (R_\mathrm{s} -r)^2 \beta}{c (1-\beta)}$ is a local
thermal timescale, $\kappa$ is the opacity, $c$ the velocity of light, and
$\beta$ the ratio of gas to total pressure.  Non-adiabatic effects become
important at locations where $\nu_\mathrm{th}$ is comparable or larger than the
forcing frequency $\bar{\sigma}$.  It can be seen in Fig.~\ref{fIM} that this is the
case for the layers in and above the inner convective shell.  At the inner edge
of this shell $\nu_\mathrm{th} \simeq \Omega_\mathrm{c}$, at the outer edge
$\nu_\mathrm{th}\simeq 5 \Omega_\mathrm{c}$.

\subsection{Turbulent viscosity in convective regions} \label{sec:turbvis} In
convective regions the equations of motion are supplemented by extra terms to
account for the occurring turbulent dissipation, as described in PS97.  For the
coefficient of turbulent viscosity $\zeta$ we adopt a simple local mixing length
approximation:  $\zeta= \alpha \, H_p\, v_\mathrm{c} $, where $H_p$ is the
pressure scale height, $\alpha=2$ is the mixing length parameter, and
$v_\mathrm{c}=\left( \frac{F}{10 \rho} \right)^{\frac{1}{3}}$ is a
characteristic convective velocity.  For high oscillation frequencies the
viscosity is reduced \cite{GK77} by a factor $ \mathrm{min}\left(1,\left(\bar{\sigma}
\tau_\mathrm{c}\right)^{-2}\right) $, where $\tau_\mathrm{c}=\alpha
H_p/v_\mathrm{c}$ is the convective timescale.  We limit $\zeta$ to be
everywhere less than $5 \times 10^{12} \mbox{ cm$^2$ s$^{-1}$}$.

In the $\mu_a$-gradient zone adjacent to the convective core, where the
Brunt-V\"ais\"al\"a fequency attains large values, the tidal response has short
wavelength and cannot be resolved on the grid when the forcing frequency
$|\bar{\sigma}|$ drops to small values.  We retain some viscosity in this region by
letting the viscosity decay outwards from the boundary of the convective core
(at $r_\mathrm{c}$) as $\zeta\propto\exp-[(r-r_\mathrm{c})/(0.1 \,H_p)]^2$.  
Note in
this respect that, although the composition gradient suppresses radial
overshooting, horizontal turbulence may be well developed in this boundary layer
between the core and envelope.  Also the oscillation amplitudes are relatively
large in this region.

\section{The torque integral: transfer of energy and angular momentum}
\label{sec:torque}
For a circular orbit the tidal potential~(\ref{eq:pot}) has no $m=0$ component.
However, since we wish to apply our results to eccentric binaries for which the
tide has an axisymmetric time dependent component we replace the factor $e^{\mathrm{i}
m(\omega t-\varphi)}$ by $e^{\mathrm{i}(\sigma t - m \varphi)}$ so that we can study 
$m=0$ forcing.  Once we have solved
Eqs.~\ref{eqco}--\ref{eqFt} for a given stellar rotation rate $\Omega_\mathrm{s}$
and a given ($l$,$m$,$\sigma$), i.e.\ for a
term
\begin{equation}
\Phi_{lm} = - f_{lm} \, r^l P_l^m (\cos
\vartheta) \, \cos (\sigma  t- \,m \varphi)
\label{eq:potc}
\end{equation}
in the forcing potential~(\ref{eq:pot}), where $f_{lm} \propto 
M_\mathrm{p}/a^{l+1}$, the rate of angular momentum exchange
with the companion's orbital motion can be calculated as an integral of the
tidal force per unit volume $\vec{F}_{lm}=-\rho \nabla \Phi_{lm}$ over the
volume of the star
\[
  \dot{H}_\mathrm{s} =
  \int_\star \vec{r} \times \vec{F}_{lm}
  \,\mathrm{d} V
\]
\[
= \mathrm{Re} \int\!\!\int\!\!\int -\frac{\partial\Phi_{lm}}{\partial\varphi} 
\rho'(r,\vartheta)\,r^2 
\,
  \sin \vartheta  \,\mathrm{d} \vartheta \,\mathrm{d} \varphi \,\mathrm{d} r \nonumber \\ 
\]
\[=  m \pi f_{lm} \int\!\!
  \int \mathrm{Im}(\rho')\, P_l^m(\cos \vartheta )\, r^{l+2} \sin \vartheta \,\mathrm{d} 
\vartheta
   \,\mathrm{d} r 
\]
\begin{equation}
\equiv m \,\mathcal{T}_{lm}
\end{equation}
where $\mathrm{Re}$ and $\mathrm{Im}$ stand for the real and imaginary part.
Since the torque and the stellar and orbital rotation vectors are all 
aligned,
only the magnitude of the transferred spin and orbital angular momentum needs 
to
be considered.    The
rate of tidal energy exchange with the (10~$\mathrm{M}_\odot$) star can be expressed as
\[
\dot{E}_\mathrm{s} = \int_\star \vec{F}_{lm} \cdot \vec{v} \,\mathrm{d} V 
\]
\begin{equation}
  =-\mathrm{Re} \int_\star \rho_0 \nabla \Phi_{lm} \cdot \vec{v}' \,\mathrm{d} V -\mathrm{Re} 
  \int_\star 
  \rho'
  \nabla \Phi_{lm} \cdot \vec{v}_\mathrm{rot} \,\mathrm{d} V.
\end{equation}
Substituting $\vec{v}' = \mathrm{i} \bar{\sigma} \vec{\xi}$ and $\vec{v}_\mathrm{rot}= 
\Omega_\mathrm{s}
r \sin \vartheta \, \hat{\vec{e}}_{\varphi}$ we find, after taking the real
part, that the second integral on the right hand side equals $m \Omega_\mathrm{s} \mathcal{T}_{lm} $.
The first integral on the right hand side can be rewritten by applying the
equation of continuity
(\ref{eqco}).  Partial integration then shows that this term is also
proportional to $\mathcal{T}_{lm} $ and equals
$\bar{\sigma} \mathcal{T}_{lm} $.  Concluding, we find for the tidally induced rate of 
respectively 
the energy and angular momentum change in the star:
\begin{equation}
  \dot{E}_\mathrm{s} = \sigma \mathcal{T}_{lm} \label{eq:en}
\end{equation}
\begin{equation}
  \dot{H}_\mathrm{s} = m \mathcal{T}_{lm} \label{eq:im}
\end{equation}
where the torque integral is given by
\begin{equation} 
  \mathcal{T}_{lm} = \pi f_{lm} \int_{-1}^{1}\int_{0}^{R_\mathrm{s}} 
  \mathrm{Im}\left(\rho'(r,\mu)\right)
  P_l^m(\mu)\, r^{l+2}  \,\mathrm{d}\mu  \,\mathrm{d} r 
  \label{eq:diss} 
\end{equation}
with $\mu=\cos \vartheta$. 
\begin{figure*} 
\includegraphics{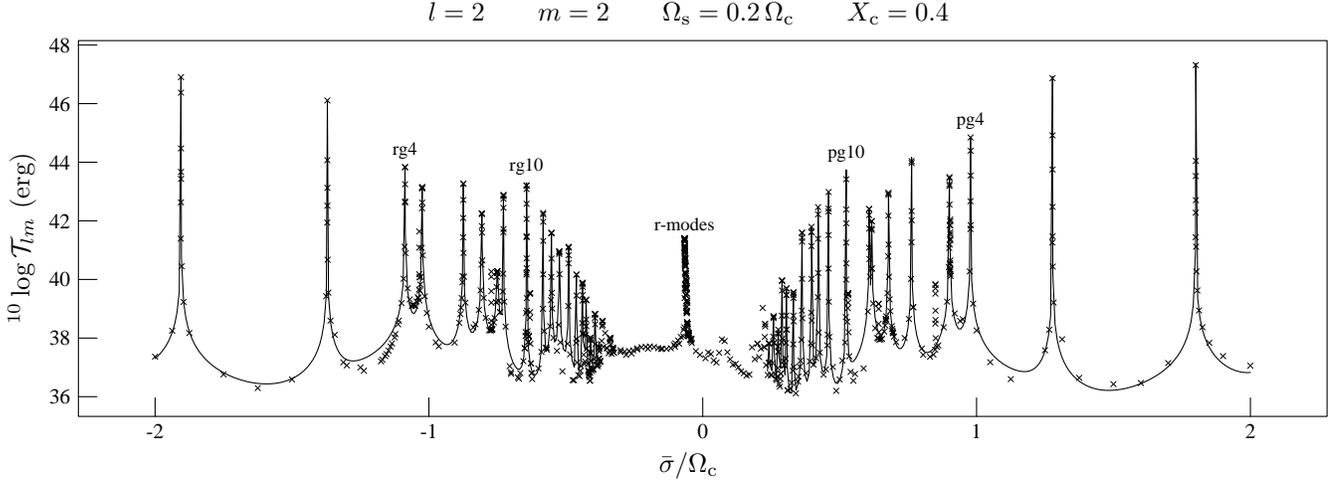} 
\caption[]{Torque integral $\mathcal{T}_{lm}$
versus forcing frequency $\bar{\sigma}$ for forcing with $l=2$ and $m=2$ on a
10~$\mathrm{M}_\odot$ star rotating at twenty percent of breakup speed.  Prograde and
retrograde g$^2_{\pm k}$-modes with $k=4$ and $10$ are labeled, as well as the 
location of
the r-modes.  Crosses denote calculated points, while the drawn continuous curve
represents a fit, see text.}  
\label{fig:m2o2} 
\end{figure*} 

\section{Numerical procedure} 
\label{sec:numproc} 
The set of partial
differential Eqs.~\ref{eqco}--\ref{eqFt} with boundary conditions
(\ref{bcs1}) and (\ref{bcs2}) is approximated by a set of finite difference
equations (FDE's) on a 2D grid in $r$ and $\vartheta$, i.e.\ on a meridional
plane through the star.  The FDE's are similar (except for the extra terms
associated with the $\mu_a$-gradient) to those described in SP97 and are solved
by the same implicit scheme as given in that paper.  We use a grid of 1200 zones
in the radial and 46 zones in the $\vartheta$ direction, analogous to the one
used in SP97.  Since the radial grid is staggered, with part of the unknown
perturbations defined on the odd and the rest on the even zones, the effective
number of radial zones is 600.

For given values of ($l$, $m$, $\sigma$) in the applied forcing potential
(\ref{eq:potc}) and for a given uniform rotation rate $\Omega_\mathrm{s}$ of the star we can
thus obtain the stellar response, i.e.\ the complex valued perturbations
$\xi_r$, $\xi_{\vartheta}$, $\xi_{\varphi}$, $\left(\frac{F'_r}{F} \right)$, 
$\left(\frac{F'_{\vartheta}}{F} \right)$, $\left(\frac{F'_{\varphi}}{F} \right)$ 
and $\left(\frac{\rho'}{\rho}\right)$, $\left(\frac{T'}{T}\right)$ on the 2D grid
in the meridional plane of the perturbed star.  By multiplying these values with
the common factor $\mathrm{e}^{\mathrm{i} \left(\sigma t- m \varphi \right)}$ we
obtain the non-adiabatic stellar response to the prescribed tidal forcing
throughout the star.

Because the 2D implicit numerical scheme involves many (complex) matrix
invertions the solution requires large computer facilities and it is essential
to use an economic method for tracing the many resonances with the free stellar
oscillation modes required to study the dynamical tide in a rotating star.  The
resonances are searched in frequency ($\bar{\sigma}$) space by by tracing the maximum
of the function \( \Psi(\bar{\sigma})=\sum _{i,j} \vec{\xi}_{i,j}\cdot
\vec{\xi}_{i,j}^{\ast} \) where $\vec{\xi}$ is the displacement vector, a $\ast$
denotes complex conjugation and $i$ and $j$ run over the radial and $\vartheta$
grid, respectively.  The maxima of $\Psi$ are searched by using a robust method
based on cubic interpolation (NAG library routine) which requires the
calculation of the first derivative 
\( {{\mathrm{d} \Psi}\over {\mathrm{d} \bar{\sigma}}}=\sum_{i,j}
\left(\vec{\xi}_{i,j} \cdot\frac{\partial\vec{\xi}_{i,j}^{\ast}}{\partial \bar{\sigma}} +
  \vec{\xi}_{i,j}^{\ast} \cdot\frac{\partial\vec{\xi}_{i,j}} {\partial \bar{\sigma}} \right).  
\) 
Fortunately,
the derivatives $\frac{\partial\vec{\xi}}{\partial \bar{\sigma}}$ and 
$\frac{\partial\vec{\xi}^{\ast}}{\partial \bar{\sigma}}$
can be obtained cheaply once we have obtained the solution of the perturbations
$\vec{\xi}$, etc.  For by differentiating Eqs.~\ref{eqco}--\ref{eqFt}
with respect to $\bar{\sigma}$ we arrive at the same equations, except that the
unknowns are now the derivatives of the perturbations (i.e.\
$\frac{\partial\vec{\xi}}{\partial \bar{\sigma}}$ instead of $\vec{\xi}$, etc.), while the right hand
sides can be expressed in terms of the solved perturbations.  The right hand
side for equation (\ref{eqmr}) for example becomes $- 2 \rho \bar{\sigma} \xi_r - 2
\mathrm{i} \rho \Omega_\mathrm{s} \sin \vartheta \, \xi_{\varphi}$.  By storing the relevant inverted
matrices during the solution procedure for the perturbations we then can simply
combine the stored matrices with the adapted right hand sides to obtain a
solution for $\frac{\partial\vec{\xi}}{\partial \bar{\sigma}}$ at almost no extra costs.
\begin{figure*} \includegraphics{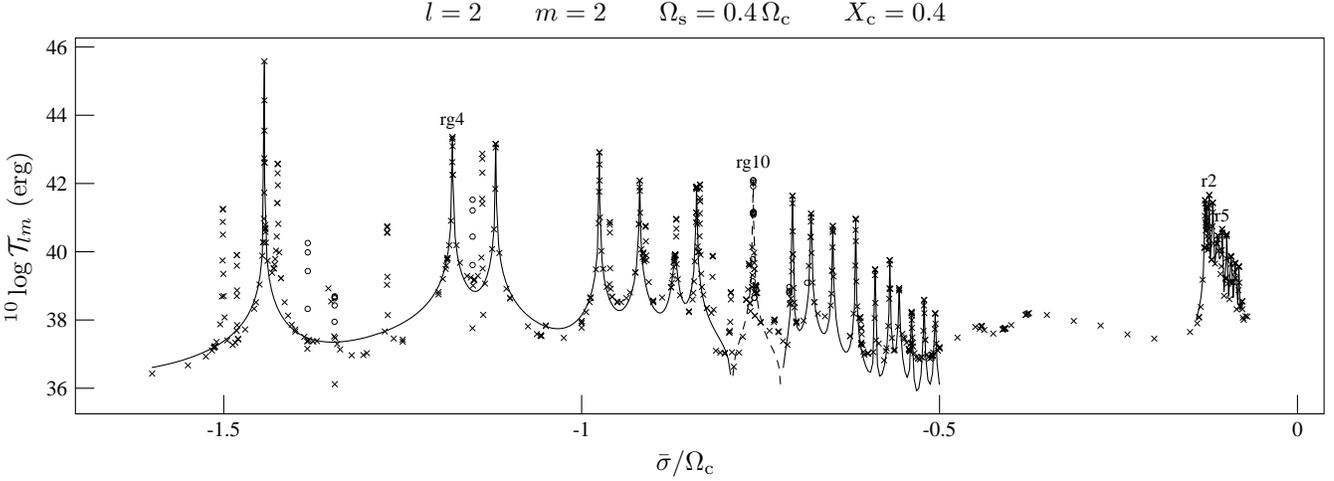} \caption[]{Torque integral $\mathcal{T}_{lm}$
versus (retrograde) forcing frequency $\bar{\sigma}$ for forcing with $l=2$ and $m=2$
on a 10~$\mathrm{M}_\odot$ star rotating at fourty percent of breakup speed.  Retrograde
g$_{-k}^2$-modes with $k=4$ and $10$ (unstable) are labeled, as well as the
r-modes with $k=2$ and $k=5$.  Crosses and circles (unstable modes) denote
calculated points.  The peaks not fitted by the continuous curve correspond to
resonances with higher spherical degrees $l=4,6,\ldots$} \label{fig:m2o4}
\end{figure*}

\section{Results} \label{sec:numres} From now on we will express all frequencies
in units of $\Omega_\mathrm{c}=1.54 \times 10^{-4} \mbox{ s$^{-1}$}$.  By varying the
forcing frequency $\bar{\sigma}=\sigma- m \Omega_\mathrm{s}$ (in the corotating frame)
for the $l=2$ terms in the forcing potential~(\ref{eq:potc}) and by calculating
the stellar response, we can search for resonances with the free stellar
oscillation modes, applying the procedure described in the previous section.  We
adopted various values for the rotation rate of the 10~$\mathrm{M}_\odot$ star:
$\Omega_\mathrm{s}=0$, 0.1, 0.2, 0.3 and 0.4.

We traced the resonances with g$^2_k$-modes (up to $k=20$) for both prograde and
retrograde g$^2$-modes.  The lower index of g denotes $k$, the number of radial
nodes in the displacement eigenvector, whereby the minus sign denotes a
retrograde mode.  Note that we characterize retrograde modes (i.e.\ modes that
propagate in the direction counter to the stellar rotation) by negative values
of the oscillation frequency $\bar{\sigma}$, i.e.\ we take $m>0$ for both prograde and
retrograde modes.  For rotating stars there is in addition a compact spectrum of
r-modes, for slightly negative frequencies.  The $l=2$ component of the tidal
potential excites r-modes with a predominant $l=3$ component.  Note that because
we consider binary systems in which both stars are aligned perpendicular to the
orbital plane the ($l=2$) forcing has only $m=0$ and $|m|=2$ components, i.e.\
$f_{21}=0$.  We can thus limit our calculations to either $|m|=2$ or $m=0$
resonances.

Note further that for all forcing frequencies $\bar{\sigma}$ we keep the factor
$f_{lm}$ in the forcing potential (\ref{eq:potc}) constant by adopting a fixed
value $a=4 R_\mathrm{s}$ and $M_\mathrm{p}=1 \mathrm{M}_\odot$.  Because we solve the linearized
problem all listed values for the tidal torque integral~(\ref{eq:diss}) can be
simply scaled to the value for a required binary configuration by
multiplying with a factor $(M_\mathrm{p}/\mathrm{M}_\odot)^2 \, (4 R_\mathrm{s}/a)^{2(l+1)}$, where
$a$ and $M_\mathrm{p}$ are the actual orbital separation and companion mass.
Unless the companion is a compact star one should of course also take into 
account the (dynamical) tide in that star to determine the tidal evolution of the 
binary system.

\subsection{Resonance fitting} Using the technique described in
Sect.~\ref{sec:numproc}, we determine the resonances with g- and r-modes and
evaluate the torque integral~(\ref{eq:diss}) from the calculated stellar
response in the meridional plane of the 10~$\mathrm{M}_\odot$ star for all applied forcing
frequencies.  Close to each resonance we fit the obtained values for the torque
integral by the resonance curve of a damped harmonic oscillator \begin{equation}
\mathcal{T}_{lm}(\bar{\sigma}) = \frac{\mathcal{T}_{lm,0}}{\left(\frac{\bar{\sigma}^2 - \bar{\sigma}_0^2}
{\bar{\sigma}_0\Delta\bar{\sigma}}\right)^2+1} \label{eq:harmosc} \end{equation} with
eigenfrequency $\bar{\sigma}_0$, resonance full width $\Delta\bar{\sigma}$ and peak value
$\mathcal{T}_{lm,0}$.  As a consequence of the Coriolis force, the stellar response to the
$l=2$ forcing contains a range of $l$-values (when we expand in spherical
harmonics).  However, we fit only the resonances with predominantly $l=2$
(g-modes) or $l=3$ (r-modes).  In between $l=2$ resonances we assume
$\mathcal{T}_{lm}(\bar{\sigma})$ can be approximated by adding the contributions of the two
adjacent $l=2$ resonances.  The detailed results of our numerical calculations
are listed in Tables~\ref{tab:m0} to~\ref{tab:rm} in the appendix.  Note that
for a non-rotating star the $l=2$ spectrum is degenerate in $m$, so that for
$\Omega_\mathrm{s}=0$ the resonance frequencies of $l=m=2$ g-modes are given by
Table~\ref{tab:m0}.

It can be seen that, as expected, for $m=0$ the g-mode resonance frequencies
$\bar{\sigma}_0$ increase with $\Omega_\mathrm{s}$, as is true for $|\bar{\sigma}_0|$ for the retrograde
$m=2$ g-modes and the r-modes.  However, the prograde $m=2$ g-mode resonances
shift to \emph{lower} oscillation frequencies as $\Omega_\mathrm{s}$ increases.  Note in this
respect that (approximate) conservation of radial vorticity, in combination with
an increasing radial component of ambient rotation $\Omega_\mathrm{s} \cos \vartheta$
towards small colatitudes, tends to give retrograde wave propagation \cite[see
e.g.]{U89}.

\subsection{Prograde and retrograde g-mode resonances} 
In Figs.~\ref{fig:m2o2}
and \ref{fig:m2o4} we show the tidal torque integral $\mathcal{T}_{lm}$ versus forcing
frequency $\bar{\sigma}$ for a 10~$\mathrm{M}_\odot$ star rotating at a rate of 0.2 and 0.4,
respectively.  Crosses and open circles correspond to numerical results, while
the continuous curve corresponds to the above mentioned fit to individual
resonances.  When the 10~$\mathrm{M}_\odot$ star rotates in the same sense as the orbital
motion of its companion, the latter cannot excite modes with frequency (in the
frame corotating with the star) less than $\bar{\sigma}=-m \Omega_\mathrm{s}$.  Nevertheless, we
have calculated the whole range $|\bar{\sigma}| \le 2$.  In Fig.~\ref{fig:m2o4}, the
circles and the dashed fit curve represent unstable modes for which the torque
integral has a reversed sign compared to $\bar{\sigma}$ (see Sect.~\ref{sec:unstab}).

The prograde g-modes shift to substantially larger oscillation frequencies
compared to the unevolved ZAMS stellar model of the same mass.  The
retrograde g-modes shift to more negative frequencies, so that for
$\Omega_\mathrm{s}=0.2$ no strong retrograde g-modes can be tidally excited, unless the
rather unlikely case of retrograde stellar rotation applies.  However, when the stellar
rotation rate is increased more and more retrograde g-modes are shifted into
the `tidal window':  up to g$^2_{-13}$ for $\Omega_\mathrm{s}=0.3$, and up to g$^2_{-10}$
for $\Omega_\mathrm{s}=0.4$. 

But note that when the binary is eccentric and the early type
star's spin is approximately synchronized at periastron, strong (retrograde)
r-modes can be tidally excited even for relatively low stellar rotation rates.

For small frequencies $|\bar{\sigma}|$ viscous damping in the $\mu_a$-gradient zone
becomes significant as the local wavelength of the response gets short and the
adopted turbulent viscosity gets large due to the comparable timescale of
`convection' and oscillations (Sect.~\ref{sec:turbvis}).  Therefore the values
for $\mathcal{T}_{lm}$ in the region between g$^2_{-20}$ and the r-modes (and similarly for
the corresponding positive frequency range) depend on the uncertain assumptions
about the viscous dissipation.  Switching off all viscous dissipation in the
$\mu_a$-gradient zone yields for $\Omega_\mathrm{s}=0.4$ in the frequency range $-0.4<
\bar{\sigma}< -0.2$ torque values about an order of magnitude smaller than shown in
Fig.~\ref{fig:m2o4}.

\subsection{Effects of rapid rotation} Although centrifugal effects are no
longer negligible for the high rotation rates ($\Omega_\mathrm{s}=0.4$) considered here
the results are still of interest, even though centrifugal distortion is
neglected, because they show the effect of strong coupling by the Coriolis
force.  Note in this respect that in the higher density interior regions the
local break up speed is substantially larger than $\Omega_\mathrm{c}$. A full calculation 
with centrifugal distortion included would require a much larger computing 
effort.

The Coriolis force gives rise to strong coupling with modes of higher spherical degrees,
so that for $l=2$ forcing resonances with $l=4,6,8,\ldots$ also become
prominent, especially when located near a $l=2$ resonance.  Because the tidal
forcing has $l=2$ symmetry the latter resonances generally dominate, but all neighbouring
resonances with $l>2$ are also excited, so that the tidal response contains
significant power over a broad range of (even) $l$-values.  Often this makes
mode identification (by means of a decomposition in Fourier-Legendre series of
the eigenfunctions, see SP97) difficult since the power in different
$l$-components are often comparable.  One should consider the excited
oscillation as a complex of coupled modes.  Since, for a given frequency the
g-modes with larger $l$ have shorter wavelength, damping is enhanced,
so that for large rotation rates the oscillation amplitude and the effective torque is reduced.
This can be observed in Tables~\ref{tab:m0}--\ref{tab:pg}, which show that the resonance area
$\pi \Delta \bar{\sigma} \mathcal{T}_{lm}$ of the $l=2$ g-modes initially increases
with $\Omega_\mathrm{s}$ but generally decreases when the rotation rate and inertial
damping becomes large.

The indirectly excited higher $l$-resonances correspond to the
peaks that have not been fitted in Fig.~\ref{fig:m2o4}.  This figure shows part
of the retrograde oscillation spectrum for $\Omega_\mathrm{s}=0.4$.  Most of the stable
unfitted peaks between g$^2_{-3}$ and g$^2_{-5}$ correspond to what are
(predominant) $l=4$ resonances, except for the two nearest resonances around g$^2_{-3}$,
which are identified as g$^8_{-12}$ and g$^{12}_{-18}$.  The unfitted peaks between
g$^2_{-6}$ and g$^2_{-10}$ are all (predominant) $l=4$ resonances, and can be
identified as g$^4_{-12}$ to g$^4_{-18}$ consecutively.  The g$^2_{-8}$ and
g$^4_{-14}$ resonances can hardly be distinguished from each other, they lie so
close together that they form a strongly coupled complex, whereby the
g$^4_{-14}$ resonance peaks one order of magnitude above the $l=2$ resonance
peak.  Another symbiosis occurs for g$^2_{-10}$ and g$^4_{-18}$, whereby,
however, the former mode appears unstable and the latter stable.  In this
frequency range the resonances with $l>4$ are heavily damped as they approach
their asymptotic low frequency regime.

For still lower frequencies in the range around $\bar{\sigma} = -0.5$ the calculated
response in between resonances is significantly stronger than the wings of the
fitted $l=2$ resonance curves due to the smeared out contributions of higher $l$
components.  At these frequencies the spectrum of $l=4$ modes starts to
become dense whereby radiative and viscous damping gets heavy due to the short wavelength of
the response.

\subsection{Unstable modes}
\label{sec:unstab}
It is now known \cite[e.g.]{DP93,GS93} that most
main sequence OB stars show unstable non-radial modes driven by the
$\kappa$-mechanism associated with the metal opacity bump around $1.5 \times
10^5$ K in the OPAL opacity tables \cite{IR96}.  Two necessary conditions for a
mode to be unstable by the $\kappa$ mechanism are that the Lagrangian pressure
perturbation $ \delta P/P$ must reach a maximum value (coming from the stellar
centre) in the driving zone and that $2 \pi/\bar{\sigma}$ is of order the thermal
timescale in that region \cite[e.g.]{DMP93}.  Obviously, when $|\bar{\sigma}|\ll
\nu_\mathrm{th} \equiv 2 \pi/ \tau_\mathrm{th}$ in the ionization region that
region is strongly non-adiabatic and no driving can occur.  The `thermal
frequency' $\nu_\mathrm{th}$ (Sect.~\ref{sec:model}) in the lower convective
shell associated with the opacity bump varies between approximately 1 and 5.
\begin{table}
   \caption[]{Unstable $m=2$ modes for $\Omega_\mathrm{s}=0.4$. The listed oscillation periods 
     $P_\mathrm{in}$ are relative to the inertial frame, a minus sign denoting retrograde 
     propagation.}
  \(
  \begin{array}{lrlr}
    \hline \rule[-1.3mm]{0mm}{4.3mm} 
    \mbox{mode} & \mathrm{Re}(\bar{\sigma}) & \multicolumn{1}{c}{\mathrm{Im}(\bar{\sigma})}
    & P_\mathrm{in} \mbox{ (d)} \\
    \hline
    \rule{0mm}{3mm}
    \mathrm{g}^{8}_{-12} & -1.3826 & -1.1 \times 10^{-5} & -0.81 \\
    \mathrm{g}^{10}_{-17} & -1.3451 & -5.3 \times 10^{-5} & -0.87 \\
    \mathrm{g}^{6}_{-13} & -1.1524 & -1.2 \times 10^{-5} & -1.34 \\
    \mathrm{g}^{2}_{-10} & -0.7602 & -1.2 \times 10^{-4} & 11.86 \\
\hline
    \mathrm{g}^{6}_{14} & 1.0580 & -4.6 \times 10^{-5} & 0.25 \\
    \mathrm{g}^{6}_{13} & 1.1261 & -4.3 \times 10^{-5} & 0.25 \\
    \mathrm{g}^{6}_{8} & 1.4610 & -1.0 \times 10^{-4} & 0.21 \\
    \mathrm{g}^{8}_{8} & 2.1887 & -2.4 \times 10^{-5} & 0.16 \\
    \hline
  \end{array}
  \label{tab:unst}
\)
\end{table}

When the stellar rotation rate is increased some inertially excited ($l>2$)
modes in our 10~$\mathrm{M}_\odot$ model appear which have a torque integral $\mathcal{T}_{lm}$ with a
sign opposite to that of $\bar{\sigma}$, which indicates instability.  This can be
checked by forcing the star with a complex frequency \cite{PAPS} and searching
for maximum amplitude in complex frequency space.  It appears that the
fundamental p$^2_0$ mode with $|\bar{\sigma}| \simeq 3.3$ is unstable for all rotation
rates $\Omega_\mathrm{s}=0$ to 0.4.  There are more unstable p-modes which we have not
studied.  Table~\ref{tab:unst} lists the unstable g-modes (prograde and retrograde) that are
found for the highest rotation rate $\Omega_\mathrm{s}=0.4$.

Only for the highest rotation rate ($\Omega_\mathrm{s}=0.4$) an unstable $l=2$ g-mode appears:
the retrograde g$^2_{-10}$ mode, with $\bar{\sigma} = -0.76 $, which is located close
to the stable g$^4_{-18}$ resonance.  The g$^2_{-10}$ mode has rather low
frequency for the $\kappa$-mechanism and is a bit peculiar in that it seems
mixed with a strong short wavelength mode trapped in the $\mu_a$ gradient zone.
The oscillation period (in inertial frame) of the unstable $l=2$ mode happens to
be quite long:  about 12 days, while the adopted stellar rotation period is 1.2
days. We also find four unstable prograde g-modes,
see Table~\ref{tab:unst}.  Our list of unstable modes is probably not exhaustive (except for
$l=2$), because we preferentially find those unstable modes which happen to be
located near $l=2$ resonances.  Unstable modes, when excited by the tidal
force, could give rise to tidal evolution counter to the normal direction of
evolution.

\subsection{r-modes} For small negative forcing frequencies around
\cite[e.g.]{PP78}:  $\bar{\sigma} \simeq - \frac{2 m \Omega_\mathrm{s}}{l (l+1)} $ strong
resonances occur with quasi-toroidal oscillation modes analogous to Rossby modes
in the earth's atmosfere \cite[e.g.]{Ped}.  For these modes the fluid elements
oscillate almost exclusively in the horizontal direction whereby the restoring
force is provided by the Coriolis force (conservation of radial component of
vorticity).  
\begin{figure} 
\includegraphics{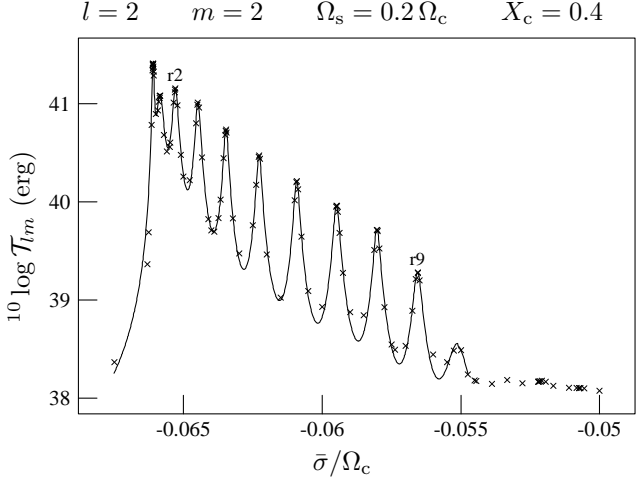} 
\caption[]{Spectrum of
r-modes:  torque integral $\mathcal{T}_{lm}$ versus forcing frequency $\bar{\sigma}$ for forcing
with $l=2$ and $m=2$ on a 10~$\mathrm{M}_\odot$ star rotating at twenty percent of breakup
speed.  Crosses denote calculated points, the drawn continuous curve represents
a fit.}  
\label{fig:rm} 
\end{figure} 
Fig.~\ref{fig:rm} shows the variation of
the torque integral $\mathcal{T}_{lm}$ as the forcing frequency runs through the r-mode
region for a star rotating with $\Omega_\mathrm{s}=0.2$.  The adopted turbulent viscosity
law in the $\mu_a$ gradient zone adjacent to the convective core causes the
r-mode resonances to become rather broadened compared to the results for ZAMS
models (SP97).  When we switch off the viscosity in the $\mu_a$-gradient zone
the peak values for the torque become one or even two orders of magnitude
higher, with a corresponding decrease of the resonance width.  However, this
will not make a large difference for the overall tidal evolution of the binary
because the total area $\simeq \pi \Delta \bar{\sigma} \mathcal{T}_{lm,0}$ of each resonance remains
approximately constant.  The resonant interaction with
r-modes gets stronger with increasing rotation rate $\Omega_\mathrm{s}$.  This can be
seen in Table~\ref{tab:rm} where the resonance area increases monotonically with $\Omega_\mathrm{s}$.

\subsection{Excitation of inertial modes in the convective core} Similar to the
results obtained in SP97 for a 20~$\mathrm{M}_\odot$ ZAMS star, we find inertial wave
oscillations in the convective core and in the convective shell region when the
forcing frequency falls in the inertial range $|\bar{\sigma}| < 2 \Omega_\mathrm{s}$.  When we
artificially cut the radiative envelope from our stellar model we find a dense
spectrum of modes in the core.  For some frequencies the resonances in the core
appear very strong.  The general amplitudes are an order of magnitude larger (or
more) compared to the full stellar model where the inertial waves can leak out
of the core.  This would suggest that the inertial modes could possibly excite
gravity waves in the envelope (SP97).  However, when we consider a model in
which the convective core is cut away, we find in this frequency region values
for the torque integral comparable to the ones calculated for the full stellar
model.  Therefore it seems the inertial core modes do not significantly
contribute to the stellar torque in this somewhat evolved stellar model.
However, the oscillations in the $\mu_a$-gradient zone outside the convective
core are poorly resolved for low frequencies, so that further studies with
significantly more meshpoints in this region seem required to draw firm
conclusions.

\section{Conclusions} We have studied the linearized non-adiabatic tidal
response of a somewhat evolved 10~$\mathrm{M}_\odot$ main sequence star to tidal forcing
with spherical harmonics degree $l=2$ and calculated the tidal exchange of
energy and angular momentum with an orbiting companion as a function of stellar
rotation rate taking the Coriolis force fully into account.  We found, as
expected, that with increasing rotation rate the inertial coupling with higher
spherical degree oscillation modes gets stronger, so that when the rotation rate
becomes comparable to or larger than the tidal oscillation frequency a
significant amount of tidal energy is transferred to these higher degree
oscillations.  Because of enhanced damping of the higher $l$ 
oscillations fast rotation reduces the resonant tidal interaction with $l=2$ g-modes.
However, this is compensated by the appearance of an additional spectrum of
($l>2$) resonances that can be excited by the $l=2$ tide.

With progressing nuclear evolution and consequent contraction of the stellar
core the frequency of the g-mode oscillations increases, so that the `gap'
between potential resonances with strong retrograde g-modes and the strong
r-modes widens.  Thereby the retrograde strong g-modes disappear from the `tidal
window' ($\bar{\sigma}_\mathrm{res} > -m \Omega_\mathrm{s}$) so that spin-down of fast spinning early type
binary stars depends on excitation of strongly damped high radial order g-modes.
This seems a not very efficient mechanism unless viscous effects (e.g.  in the
$\mu_a$-gradient zone outside the convective core) are significant or perhaps
driving by inertial resonances in the convective core occurs.  But, when in an
eccentric binary system, the early type star is nearly corotating at periastron
r-mode resonances are very efficient in spinning the star further down to
(pseudo)corotation.  We expect a balance between spin-down by r-modes and
spin-up by prograde g-modes excited by the high frequency components of the
tidal forces near periastron.  We intend to study this in a following paper, using
the torque values determined here.

\begin{acknowledgements}
  This work was sponsored by the Stichting Nationale
  Computerfaciliteiten (National Computing Facilities Foundation, NCF)
  for the use of supercomputing facilities, with financial support
  from the Netherlands Organization for Scientific Research (NWO).
\end{acknowledgements}

\bibliographystyle{aamar}
\bibliography{mnemomar,astromarbib}

\onecolumn
\appendix
\section{Tables}
\label{sec:tbl}
\begin{table*}[h!]
  \caption[]{Fitting parameters for g$^2_k$-mode resonances with an $l=2$, $m=0$
tidal potential in a 10~$\mathrm{M}_\odot$ star with $X_c=0.4$. All tabulated values for the
torque integral have been obtained for a fixed orbital separation.}
  \( 
  \begin{array}{l@{\quad}lrl@{\quad}lrl@{\quad}lrl}
    \hline &
    \multicolumn{3}{l}{\Omega_\mathrm{s}=0.0\,\Omega_\mathrm{c}} &
    \multicolumn{3}{l}{\Omega_\mathrm{s}=0.2\,\Omega_\mathrm{c}} &
    \multicolumn{3}{l}{\Omega_\mathrm{s}=0.4\,\Omega_\mathrm{c}} \\
    k &
    \multicolumn{1}{l}{\bar{\sigma}_0/\Omega_\mathrm{c}} &
    \multicolumn{1}{l}{\mathcal{T}_{lm,0} \mbox{ (erg)}} &
    \multicolumn{1}{l}{\Delta\bar{\sigma}/\Omega_\mathrm{c}} &
    \multicolumn{1}{l}{\bar{\sigma}_0/\Omega_\mathrm{c}} &
    \multicolumn{1}{l}{\mathcal{T}_{lm,0} \mbox{ (erg)}} &
    \multicolumn{1}{l}{\Delta\bar{\sigma}/\Omega_\mathrm{c}} &
    \multicolumn{1}{l}{\bar{\sigma}_0/\Omega_\mathrm{c}} &
    \multicolumn{1}{l}{\mathcal{T}_{lm,0} \mbox{ (erg)}} &
    \multicolumn{1}{l}{\Delta\bar{\sigma}/\Omega_\mathrm{c}} \\
    \hline \rule{0mm}{3.5mm}
    1  &  2.2186 &  2.94\times 10^{45} & 1.33\times 10^{-5} & 2.2344 &  
2.95\times 10^{45} & 1.32\times 10^{-5} & 2.2804 & 3.08\times 10^{45} & 
1.28\times 10^{-5} \\
    2  &  1.8489 &  3.35\times 10^{45} & 1.34\times 10^{-5} & 1.8676 &  
2.72\times 10^{45} & 1.64\times 10^{-5} & 1.9214 & 3.19\times 10^{45} & 
1.38\times 10^{-5} \\
    3  &  1.3162 &  9.00\times 10^{44} & 1.18\times 10^{-5} & 1.3427 &  
8.06\times 10^{44} & 1.39\times 10^{-5} & 1.4196 & 7.38\times 10^{43} & 
1.01\times 10^{-4} \\
    4  &  1.0203 &  9.28\times 10^{42} & 2.70\times 10^{-4} & 1.0538 &  
6.81\times 10^{42} & 3.63\times 10^{-4} & 1.1465 & 9.75\times 10^{42} & 
1.87\times 10^{-4} \\
    5  &  0.9510 &  5.08\times 10^{41} & 1.37\times 10^{-3} & 0.9862 &  
5.30\times 10^{41} & 1.22\times 10^{-3} & 1.0806 & 6.14\times 10^{41} & 
1.05\times 10^{-3} \\
    6  &  0.8037 &  1.42\times 10^{42} & 4.12\times 10^{-4} & 0.8454 &  
1.46\times 10^{42} & 3.89\times 10^{-4} & 0.9532 & 1.76\times 10^{42} & 
3.03\times 10^{-4} \\
    7  &  0.7258 &  9.82\times 10^{40} & 1.64\times 10^{-3} & 0.7711 &  
1.04\times 10^{41} & 1.44\times 10^{-3} & 0.8852 & 1.04\times 10^{41} & 
1.15\times 10^{-3} \\
    8  &  0.6637 &  2.74\times 10^{39} & 1.03\times 10^{-2} & 0.7119 &  
3.09\times 10^{39} & 8.43\times 10^{-3} & 0.8338 & 1.26\times 10^{40} & 
1.68\times 10^{-3} \\
    9  &  0.6503 &  2.07\times 10^{41} & 7.00\times 10^{-4} & 0.7006 &  
1.73\times 10^{41} & 7.90\times 10^{-4} & 0.8217 & 1.85\times 10^{41} & 
5.76\times 10^{-4} \\
    10 &  0.5628 &  1.62\times 10^{42} & 5.22\times 10^{-5} & 0.6198 &  
1.57\times 10^{42} & 4.88\times 10^{-5} & 0.7519 & 2.41\times 10^{41} & 
2.49\times 10^{-4} \\
    11 &  0.4976 &  1.86\times 10^{41} & 1.46\times 10^{-4} & 0.5608 &  
2.09\times 10^{41} & 1.16\times 10^{-4} & 0.6988 & 2.93\times 10^{40} & 
6.63\times 10^{-4} \\
    12 &  0.4619 &  2.72\times 10^{40} & 6.96\times 10^{-4} & 0.5286 &  
3.20\times 10^{40} & 5.07\times 10^{-4} & 0.6693 & 1.98\times 10^{40} & 
4.92\times 10^{-4} \\
    13 &  0.4345 &  2.43\times 10^{40} & 2.85\times 10^{-4} & 0.5048 &  
2.11\times 10^{40} & 2.58\times 10^{-4} & 0.6471 & 5.06\times 10^{39} & 
4.43\times 10^{-4} \\
    14 &  0.3983 &  3.18\times 10^{40} & 1.76\times 10^{-4} & 0.4734 &  
2.58\times 10^{40} & 1.70\times 10^{-4} & 0.6183 & 7.39\times 10^{39} & 
2.33\times 10^{-4} \\
    15 &  0.3662 &  8.01\times 10^{38} & 2.84\times 10^{-4} & 0.4459 &  
6.71\times 10^{38} & 2.04\times 10^{-4} & 0.5917 & 1.99\times 10^{37} & 
3.38\times 10^{-4} \\
    16 &  0.3404 &  4.29\times 10^{38} & 8.72\times 10^{-4} & 0.4240 &  
4.78\times 10^{38} & 5.53\times 10^{-4} & 0.5700 & 2.14\times 10^{38} & 
4.55\times 10^{-4} \\
    17 &  0.3248 &  2.05\times 10^{38} & 2.04\times 10^{-3} & 0.4109 &  
2.79\times 10^{38} & 1.21\times 10^{-3} & 0.5561 & 2.22\times 10^{38} & 
9.41\times 10^{-4} \\
    18 &  0.3086 &  1.02\times 10^{37} & 1.60\times 10^{-3} & 0.3977 &  
1.45\times 10^{37} & 1.00\times 10^{-3} & 0.5422 & 2.87\times 10^{37} & 
1.34\times 10^{-3} \\
    19 &  0.2904 &  8.10\times 10^{37} & 1.47\times 10^{-3} & 0.3827 &  
9.18\times 10^{37} & 9.17\times 10^{-4} & 0.5262 & 3.99\times 10^{37} & 
7.42\times 10^{-4} \\
    20 &  0.2734 &  5.86\times 10^{36} & 2.82\times 10^{-3} & 0.3689 &  
1.05\times 10^{37} & 1.95\times 10^{-3} & 0.5105 & 2.24\times 10^{37} & 
1.04\times 10^{-3} \\
    \hline
  \end{array}
  \label{tab:m0}
  \)
\end{table*}
\begin{table*}[h!]

  \caption[]{Fitting parameters for {\it retrograde} g$^2_k$-mode resonances with 
an
$l=2$, $m=2$ tidal potential in a 10 $\mathrm{M}_\odot$ star with $X_c=0.4$. All tabulated 
values for the torque integral
have been obtained for a fixed orbital separation.}

  \( 
  \begin{array}{l@{\quad}lrl@{\quad}lrl@{\quad}lrl}
    \hline &
    \multicolumn{3}{l}{\Omega_\mathrm{s}=0.2\,\Omega_\mathrm{c}} &
    \multicolumn{3}{l}{\Omega_\mathrm{s}=0.3\,\Omega_\mathrm{c}} &
    \multicolumn{3}{l}{\Omega_\mathrm{s}=0.4\,\Omega_\mathrm{c}} \\
    k &
    \multicolumn{1}{l}{\bar{\sigma}_0/\Omega_\mathrm{c}} &
    \multicolumn{1}{l}{\mathcal{T}_{lm,0} \mbox{ (erg)}} &
    \multicolumn{1}{l}{\Delta\bar{\sigma}/\Omega_\mathrm{c}} &
    \multicolumn{1}{l}{\bar{\sigma}_0/\Omega_\mathrm{c}} &
    \multicolumn{1}{l}{\mathcal{T}_{lm,0} \mbox{ (erg)}} &
    \multicolumn{1}{l}{\Delta\bar{\sigma}/\Omega_\mathrm{c}} &
    \multicolumn{1}{l}{\bar{\sigma}_0/\Omega_\mathrm{c}} &
    \multicolumn{1}{l}{\mathcal{T}_{lm,0} \mbox{ (erg)}} &
    \multicolumn{1}{l}{\Delta\bar{\sigma}/\Omega_\mathrm{c}} \\
    \hline \rule{0mm}{3.5mm}
    1  & -2.2734 & -5.59\times 10^{46} &  3.00\times 10^{-6} & -2.3042 & 
-4.05\times 10^{46} & 3.39\times 10^{-6} & -2.3376 & -2.91\times 10^{46} & 
3.82\times 10^{-6} \\
    2  & -1.9055 & -8.22\times 10^{46} &  2.75\times 10^{-6} & -1.9368 & 
-6.93\times 10^{46} & 2.77\times 10^{-6} & -1.9700 & -5.54\times 10^{46} & 
3.05\times 10^{-6} \\
    3  & -1.3709 & -1.29\times 10^{46} &  3.80\times 10^{-6} & -1.4047 & 
-9.02\times 10^{45} & 4.58\times 10^{-6} & -1.4431 & -3.86\times 10^{45} & 
8.73\times 10^{-6} \\
    4  & -1.0872 & -6.92\times 10^{43} &  1.41\times 10^{-4} & -1.1306 & 
-3.91\times 10^{43} & 1.95\times 10^{-4} & -1.1804 & -2.28\times 10^{43} & 
2.54\times 10^{-4} \\
    5  & -1.0246 & -1.42\times 10^{43} &  2.97\times 10^{-4} & -1.0697 & 
-1.41\times 10^{43} & 2.77\times 10^{-4} & -1.1200 & -1.45\times 10^{43} & 
2.70\times 10^{-4} \\
    6  & -0.8745 & -1.88\times 10^{43} &  1.31\times 10^{-4} & -0.9216 & 
-1.28\times 10^{43} & 1.56\times 10^{-4} & -0.9752 & -8.23\times 10^{42} & 
1.89\times 10^{-4} \\
    7  & -0.8064 & -1.79\times 10^{42} &  4.57\times 10^{-4} & -0.8592 & 
-1.40\times 10^{42} & 5.35\times 10^{-4} & -0.9189 & -1.22\times 10^{42} & 
4.96\times 10^{-4} \\
    8  & -0.7501 & -1.93\times 10^{40} &  3.07\times 10^{-3} & -0.8065 & 
-1.17\times 10^{40} & 3.51\times 10^{-3} & -0.8693 & -8.35\times 10^{39} & 
3.84\times 10^{-3} \\
    9  & -0.7277 & -7.76\times 10^{42} &  9.15\times 10^{-5} & -0.7794 & 
-7.71\times 10^{42} & 7.48\times 10^{-5} & -0.8395 & -7.92\times 10^{41} & 
4.30\times 10^{-4} \\
    10 & -0.6434 & -2.01\times 10^{43} &  1.92\times 10^{-5} & -0.6985 & 
-9.26\times 10^{42} & 3.33\times 10^{-5} & -0.7602 &  1.27\times 10^{42} & 
1.97\times 10^{-4} \\
    11 & -0.5827 & -1.88\times 10^{42} &  5.97\times 10^{-5} & -0.6417 & 
-7.80\times 10^{41} & 1.05\times 10^{-4} & -0.7054 & -4.40\times 10^{41} & 
1.25\times 10^{-4} \\
    12 & -0.5525 & -3.89\times 10^{41} &  2.51\times 10^{-4} & -0.6143 & 
-3.04\times 10^{41} & 2.75\times 10^{-4} & -0.6795 & -1.30\times 10^{41} & 
4.90\times 10^{-4} \\
    13 & -0.5238 & -8.97\times 10^{40} &  5.71\times 10^{-4} & -0.5849 & 
-2.58\times 10^{41} & 1.31\times 10^{-4} & -0.6491 & -5.72\times 10^{40} & 
4.04\times 10^{-4} \\
    14 & -0.4899 & -1.29\times 10^{41} &  2.63\times 10^{-4} & -0.5525 & 
-1.73\times 10^{41} & 1.60\times 10^{-4} & -0.6170 & -9.25\times 10^{40} & 
1.90\times 10^{-4} \\
    15 & -0.4612 & -1.47\times 10^{40} &  1.44\times 10^{-4} & -0.5253 & 
-6.51\times 10^{39} & 2.64\times 10^{-4} & -0.5900 & -3.05\times 10^{39} & 
2.33\times 10^{-4} \\
    16 & -0.4390 & -7.70\times 10^{39} &  4.13\times 10^{-4} & -0.5046 & 
-6.79\times 10^{39} & 4.52\times 10^{-4} & -0.5695 & -5.67\times 10^{39} & 
4.15\times 10^{-4} \\
    17 & -0.4259 & -2.04\times 10^{39} &  6.86\times 10^{-4} & -0.4919 & 
-1.24\times 10^{39} & 8.12\times 10^{-4} & -0.5564 & -8.61\times 10^{38} & 
9.36\times 10^{-4} \\
    18 & -0.4101 & -1.02\times 10^{38} &  6.33\times 10^{-4} & -0.4757 & 
-1.15\times 10^{38} & 8.91\times 10^{-4} & -0.5391 & -1.80\times 10^{38} & 
6.93\times 10^{-4} \\
    19 & -0.3935 & -6.68\times 10^{38} &  9.02\times 10^{-4} & -0.4592 & 
-6.61\times 10^{38} & 6.68\times 10^{-4} & -0.5216 & -3.90\times 10^{38} & 
6.60\times 10^{-4} \\
    20 & -0.3787 & -4.97\times 10^{37} &  2.03\times 10^{-3} & -0.4445 & 
-7.44\times 10^{37} & 1.29\times 10^{-3} & -0.5060 & -1.57\times 10^{38} & 
9.52\times 10^{-4} \\
    \hline
  \end{array}
  \label{tab:rg}
  \)
\end{table*}
\begin{table*}[h!]
  \caption[]{Fitting parameters for {\it prograde} g$^2_k$-mode resonances with 
an
$l=2$, $m=2$ tidal potential in a 10~$\mathrm{M}_\odot$ star with $X_c=0.4$. All tabulated 
values for the torque integral
have been obtained for a fixed orbital separation.}

  \(
  \begin{array}{l@{\quad}lrl@{\quad}lrl@{\quad}lrl}
    \hline &
    \multicolumn{3}{l}{\Omega_\mathrm{s}=0.2\,\Omega_\mathrm{c}} &
    \multicolumn{3}{l}{\Omega_\mathrm{s}=0.3\,\Omega_\mathrm{c}} &
    \multicolumn{3}{l}{\Omega_\mathrm{s}=0.4\,\Omega_\mathrm{c}} \\
    k &
    \multicolumn{1}{l}{\bar{\sigma}_0/\Omega_\mathrm{c}} &
    \multicolumn{1}{l}{\mathcal{T}_{lm,0} \mbox{ (erg)}} &
    \multicolumn{1}{l}{\Delta\bar{\sigma}/\Omega_\mathrm{c}} &
    \multicolumn{1}{l}{\bar{\sigma}_0/\Omega_\mathrm{c}} &
    \multicolumn{1}{l}{\mathcal{T}_{lm,0} \mbox{ (erg)}} &
    \multicolumn{1}{l}{\Delta\bar{\sigma}/\Omega_\mathrm{c}} &
    \multicolumn{1}{l}{\bar{\sigma}_0/\Omega_\mathrm{c}} &
    \multicolumn{1}{l}{\mathcal{T}_{lm,0} \mbox{ (erg)}} &
    \multicolumn{1}{l}{\Delta\bar{\sigma}/\Omega_\mathrm{c}} \\
    \hline \rule{0mm}{3.5mm}
    1  & 2.1724 &  1.92\times 10^{47} & 1.55\times 10^{-6} & 2.1519 &  
2.38\times 10^{47} & 1.43\times 10^{-6} & 2.1330 &  2.98\times 10^{47} & 
1.26\times 10^{-6} \\
    2  & 1.8006 &  2.31\times 10^{47} & 1.39\times 10^{-6} &     1.7781  &       
1.49\times 10^{47} &
 3.53\times 10^{-6} & 1.7587 &  6.71\times 10^{46} & 5.67\times 10^{-6} \\
    3  & 1.2773 &  8.31\times 10^{46} & 1.11\times 10^{-6} & 1.2623 &  
6.55\times 10^{46} & 1.61\times 10^{-6} & 1.2498 &  5.15\times 10^{45} & 
2.28\times 10^{-5} \\
    4  & 0.9780 &  7.33\times 10^{44} & 2.94\times 10^{-5} & 0.9638 &  
1.37\times 10^{45} & 1.87\times 10^{-5} & 0.9532 &  2.65\times 10^{45} & 
1.13\times 10^{-5} \\
    5  & 0.9011 &  3.11\times 10^{43} & 1.33\times 10^{-4} & 0.8836 &  
3.26\times 10^{43} & 1.23\times 10^{-4} & 0.8692 &  6.30\times 10^{43} & 
7.10\times 10^{-5} \\
    6  & 0.7629 &  1.46\times 10^{44} & 3.29\times 10^{-5} & 0.7508 &  
2.57\times 10^{44} & 2.17\times 10^{-5} & 0.7425 &  2.56\times 10^{44} & 
2.56\times 10^{-5} \\
    7  & 0.6783 &  9.16\times 10^{42} & 1.27\times 10^{-4} & 0.6637 &  
1.61\times 10^{43} & 8.03\times 10^{-5} & 0.6534 &  2.18\times 10^{43} & 
6.48\times 10^{-5} \\
    8  & 0.6171 &  1.00\times 10^{42} & 5.41\times 10^{-4} & 0.6044 &  
3.29\times 10^{42} & 2.64\times 10^{-4} & 0.5959 &  6.81\times 10^{42} & 
1.62\times 10^{-4} \\
    9  & 0.6078 &  2.55\times 10^{42} & 2.76\times 10^{-4} & 0.5948 &  
1.66\times 10^{42} & 2.75\times 10^{-4} & 0.5853 &  7.28\times 10^{41} & 
3.55\times 10^{-4} \\
    10 & 0.5242 &  5.54\times 10^{43} & 1.11\times 10^{-5} & 0.5152 &  
3.09\times 10^{43} & 2.21\times 10^{-5} & 0.5103 &  1.05\times 10^{43} & 
7.07\times 10^{-5} \\
    11 & 0.4596 &  9.80\times 10^{42} & 2.08\times 10^{-5} & 0.4517 &  
3.54\times 10^{42} & 6.27\times 10^{-5} & 0.4476 &  4.44\times 10^{42} & 
5.33\times 10^{-5} \\
    12 & 0.4213 &  2.92\times 10^{42} & 4.04\times 10^{-5} & 0.4125 &  
5.07\times 10^{42} & 2.35\times 10^{-5} & 0.4077 &  6.97\times 10^{42} & 
1.70\times 10^{-5} \\
    13 & 0.3971 &  6.25\times 10^{41} & 5.02\times 10^{-5} & 0.3896 &  
2.15\times 10^{41} & 1.19\times 10^{-4} & 0.3855 &  2.86\times 10^{41} & 
6.81\times 10^{-5} \\
    14 & 0.3627 &  4.18\times 10^{41} & 6.77\times 10^{-5} & 0.3565 &  
2.87\times 10^{41} & 8.87\times 10^{-5} & 0.3536 &  1.76\times 10^{41} & 
1.30\times 10^{-4} \\
    15 & 0.3313 &  3.67\times 10^{39} & 1.56\times 10^{-4} & 0.3256 &  
1.99\times 10^{39} & 1.69\times 10^{-4} & 0.3232 &  1.54\times 10^{39} & 
1.24\times 10^{-4} \\
    16 & 0.3057 &  5.00\times 10^{39} & 2.52\times 10^{-4} & 0.3004 &  
3.85\times 10^{39} & 2.43\times 10^{-4} & 0.2981 &  3.26\times 10^{39} & 
2.33\times 10^{-4} \\
    17 & 0.2896 &  9.20\times 10^{39} & 2.98\times 10^{-4} & 0.2840 &  
9.48\times 10^{39} & 2.83\times 10^{-4} & 0.2813 &  9.43\times 10^{39} & 
2.65\times 10^{-4} \\
    18 & 0.2758 &  2.06\times 10^{38} & 6.43\times 10^{-4} & 0.2711 &  
3.34\times 10^{38} & 6.40\times 10^{-4} & 0.2690 &  6.06\times 10^{38} & 
5.35\times 10^{-4} \\
    19 & 0.2586 &  5.45\times 10^{38} & 1.08\times 10^{-3} & 0.2545 &  
5.09\times 10^{38} & 1.01\times 10^{-3} & 0.2529 &  4.88\times 10^{38} & 
9.27\times 10^{-4} \\
    20 & 0.2421 &  6.70\times 10^{37} & 1.66\times 10^{-3} & 0.2384 &  
8.80\times 10^{37} & 1.47\times 10^{-3} & 0.2370 &  1.05\times 10^{38} & 
1.28\times 10^{-3} \\
    \hline
  \end{array}
  \label{tab:pg}
  \)
\end{table*}
\begin{table*}[h!]
  \caption[]{Fitting parameters for r$^3_k$-mode resonances with an $l=2$, $m=2$
    tidal potential in a 10 $\mathrm{M}_\odot$ star with $X_c=0.4$. All tabulated values 
for the torque integral
have been obtained for a fixed orbital separation.}
  \(
  \begin{array}{l@{\quad}lrl@{\quad}lrl}
    \hline &
    \multicolumn{3}{l}{\Omega_\mathrm{s}=0.1\,\Omega_\mathrm{c}} &
    \multicolumn{3}{l}{\Omega_\mathrm{s}=0.2\,\Omega_\mathrm{c}} \\
    k &
    \multicolumn{1}{l}{\bar{\sigma}_0/\Omega_\mathrm{c}} &
    \multicolumn{1}{l}{\mathcal{T}_{lm,0} \mbox{ (erg)}} &
    \multicolumn{1}{l}{\Delta\bar{\sigma}/\Omega_\mathrm{c}} &
    \multicolumn{1}{l}{\bar{\sigma}_0/\Omega_\mathrm{c}} &
    \multicolumn{1}{l}{\mathcal{T}_{lm,0} \mbox{ (erg)}} &
    \multicolumn{1}{l}{\Delta\bar{\sigma}/\Omega_\mathrm{c}} \\
    \hline \rule{0mm}{3.5mm}
    0  & -3.3237\times 10^{-2} & -5.63\times 10^{41} & 5.71\times 10^{-6} & 
-6.6098\times 10^{-2} & -2.57\times 10^{41} &  7.35\times 10^{-5} \\
    1  & -3.3191\times 10^{-2} & -7.69\times 10^{40} & 4.80\times 10^{-5} & 
-6.5845\times 10^{-2} & -1.20\times 10^{41} &  2.40\times 10^{-4} \\
    2  & -3.3133\times 10^{-2} & -1.04\times 10^{41} & 3.61\times 10^{-5} & 
-6.5292\times 10^{-2} & -1.41\times 10^{41} &  2.03\times 10^{-4} \\
    3  & -3.3031\times 10^{-2} & -8.46\times 10^{40} & 3.03\times 10^{-5} & 
-6.4479\times 10^{-2} & -1.03\times 10^{41} &  1.87\times 10^{-4} \\
    4  & -3.2896\times 10^{-2} & -4.67\times 10^{40} & 3.14\times 10^{-5} & 
-6.3467\times 10^{-2} & -5.42\times 10^{40} &  1.85\times 10^{-4} \\
    5  & -3.2730\times 10^{-2} & -2.43\times 10^{40} & 3.46\times 10^{-5} & 
-6.2277\times 10^{-2} & -2.96\times 10^{40} &  2.00\times 10^{-4} \\
    6  & -3.2534\times 10^{-2} & -1.27\times 10^{40} & 3.97\times 10^{-5} & 
-6.0922\times 10^{-2} & -1.62\times 10^{40} &  2.11\times 10^{-4} \\
    7  & -3.2315\times 10^{-2} & -7.93\times 10^{39} & 4.26\times 10^{-5} & 
-5.9485\times 10^{-2} & -9.14\times 10^{39} &  2.38\times 10^{-4} \\
    8  & -3.2084\times 10^{-2} & -4.02\times 10^{39} & 4.63\times 10^{-5} & 
-5.8029\times 10^{-2} & -5.19\times 10^{39} &  2.51\times 10^{-4} \\
    9  & -3.1842\times 10^{-2} & -1.48\times 10^{39} & 5.51\times 10^{-5} & 
-5.6565\times 10^{-2} & -1.89\times 10^{39} &  3.12\times 10^{-4} \\
    10 & -3.1581\times 10^{-2} & -2.24\times 10^{38} & 8.18\times 10^{-5} & 
-5.5121\times 10^{-2} & -3.39\times 10^{38} &  7.66\times 10^{-4} \\
    \hline &
    \multicolumn{3}{l}{\Omega_\mathrm{s}=0.3\,\Omega_\mathrm{c}} &
    \multicolumn{3}{l}{\Omega_\mathrm{s}=0.4\,\Omega_\mathrm{c}} \\
    k &
    \multicolumn{1}{l}{\bar{\sigma}_0/\Omega_\mathrm{c}} &
    \multicolumn{1}{l}{\mathcal{T}_{lm,0} \mbox{ (erg)}} &
    \multicolumn{1}{l}{\Delta\bar{\sigma}/\Omega_\mathrm{c}} &
    \multicolumn{1}{l}{\bar{\sigma}_0/\Omega_\mathrm{c}} &
    \multicolumn{1}{l}{\mathcal{T}_{lm,0} \mbox{ (erg)}} &
    \multicolumn{1}{l}{\Delta\bar{\sigma}/\Omega_\mathrm{c}} \\
    \hline \rule{0mm}{3.5mm}
    0  & -9.8117\times 10^{-2} & -1.74\times 10^{41} & 3.87\times 10^{-4} & 
-1.2898\times 10^{-1} & -3.26\times 10^{41} & 4.25\times 10^{-4} \\
    1  & -9.7284\times 10^{-2} & -1.12\times 10^{41} & 9.12\times 10^{-4} & 
-1.2685\times 10^{-1} & -2.23\times 10^{41} & 6.54\times 10^{-4} \\
    2  & -9.5530\times 10^{-2} & -1.95\times 10^{41} & 3.78\times 10^{-4} & 
-1.2328\times 10^{-1} & -4.60\times 10^{41} & 3.18\times 10^{-4} \\
    3  & -9.3144\times 10^{-2} & -1.51\times 10^{41} & 3.93\times 10^{-4} & 
-1.1833\times 10^{-1} & -2.73\times 10^{41} & 4.26\times 10^{-4} \\
    4  & -9.0057\times 10^{-2} & -4.14\times 10^{40} & 7.76\times 10^{-4} & 
-1.1145\times 10^{-1} & -3.08\times 10^{40} & 2.42\times 10^{-3} \\
    5  & -8.6610\times 10^{-2} & -2.93\times 10^{40} & 4.95\times 10^{-4} & 
-1.0515\times 10^{-1} & -4.60\times 10^{40} & 5.49\times 10^{-4} \\
    6  & -8.2982\times 10^{-2} & -2.23\times 10^{40} & 4.14\times 10^{-4} & 
-9.8842\times 10^{-2} & -3.21\times 10^{40} & 5.26\times 10^{-4} \\
    7  & -7.9327\times 10^{-2} & -9.10\times 10^{39} & 5.65\times 10^{-4} & 
-9.2639\times 10^{-2} & -7.71\times 10^{39} & 1.12\times 10^{-3} \\
    8  & -7.5855\times 10^{-2} & -5.14\times 10^{39} & 6.24\times 10^{-4} & 
-8.6945\times 10^{-2} & -5.05\times 10^{39} & 1.08\times 10^{-3} \\
    9  & -7.2437\times 10^{-2} & -2.03\times 10^{39} & 7.43\times 10^{-4} & 
-8.1781\times 10^{-2} & -3.74\times 10^{39} & 7.49\times 10^{-4} \\
    10 & -6.9712\times 10^{-2} & -2.28\times 10^{38} & 3.98\times 10^{-3} & 
-7.7343\times 10^{-2} & -3.83\times 10^{38} & 1.02\times 10^{-3} \\
    \hline
  \end{array}
  \label{tab:rm}
  \)
\end{table*}

\end{document}